\documentclass[preprint]{aastex}
\title{A decline and fall in the future of Italian Astronomy?}
\author{Angelo Antonelli$^1$, Vincenzo Antonuccio-Delogu$^2$, 
  Andrea Baruffolo$^3$, Stefano Benetti$^3$, 
  Simone Bianchi$^4$, Andrea Biviano$^5$,
  Annalisa Bonafede$^6$, Marco Bondi$^6$,
  Stefano Borgani$^{7,5,8}$, Angela Bragaglia$^9$, 
  Massimo Brescia$^{10}$, John Robert Brucato$^4$, 
  Gianfranco Brunetti$^6$, Riccardo Brunino$^{11}$, 
  Michele Cantiello$^{11}$, Viviana Casasola$^6$, 
  Rossella Cassano$^6$, Alberto Cellino$^{13}$, 
  Gabriele Cescutti$^7$, Andrea Cimatti$^{14}$, 
  Andrea Comastri$^9$, Edvige Corbelli$^4$, 
  Giovanni Cresci$^4$, Serena Criscuoli$^1$,
  Stefano Cristiani$^{5,8}$, Guido Cupani$^5$,
  Sabrina De Grandi$^{15}$, Valerio D'Elia$^1$, 
  Melania Del Santo$^{16}$, Gabriella De Lucia$^5$, 
  Silvano Desidera$^3$, Marcella Di Criscienzo$^1$, 
  Valentina D'Odorico$^5$, Elisabetta Dotto$^1$, 
  Fabio Fontanot$^5$, Mario Gai$^{13}$, 
  Simona Gallerani$^1$, Stefano Gallozzi$^1$, 
  Bianca Garilli$^{17}$, Isabella Gioia$^6$, 
  Marisa Girardi$^{7,5}$, Myriam Gitti$^9$, 
  Gianluigi Granato$^5$, Raffaele Gratton$^3$, 
  Andrea Grazian$^1$, Carlotta Gruppioni$^9$, 
  Leslie Hunt$^4$, Giuseppe Leto$^2$, 
  Gianluca Israel$^1$, Manuela Magliocchetti$^{18}$, 
  Laura Magrini$^4$, Gabriele Mainetti$^{19}$, 
  Filippo Mannucci$^4$, Alessandro Marconi$^{20}$, 
  Martino Marelli$^{17,21}$, Michele Maris$^5$, 
  Francesca Matteucci$^{7,5}$, Massimo Meneghetti$^9$, 
  Aniello Mennella$^{22}$, Amata Mercurio$^{10}$, 
  Silvano Molendi$^{17}$, Pierluigi Monaco$^{7,5}$, 
  Alessia Moretti$^3$, Giuseppe Murante$^{13}$, 
  Fabrizio Nicastro$^1$, Marina Orio$^3$, 
  Adamantia Paizis$^{17}$, Francesca Panessa$^{16}$, 
  Fabio Pasian$^5$, Laura Pentericci$^1$, 
  Lucia Pozzetti$^9$, Mariachiara Rossetti$^{17}$, 
  Joana S. Santos$^5$, Alexandro Saro$^{7,8}$, 
  Raffaella Schneider$^4$, Laura Silva$^5$, 
  Roberto Silvotti$^{13}$, Richard Smart$^{13}$, 
  Andrea Tiengo$^{17}$, Luca Tornatore$^{7,5}$, 
  Paolo Tozzi$^5$, Edoardo Trussoni$^{13}$, 
  Tiziano Valentinuzzi$^3$, Eros Vanzella$^5$, 
  Franco Vazza$^6$, Alberto Vecchiato$^{13}$, 
  Tiziana Venturi$^6$, Giacomo Vianello$^{23}$, 
  Matteo Viel$^5$, Alvaro Villalobos$^5$, 
  Valentina Viotto$^{3,19}$, Benedetta Vulcani$^{19,3}$\\ 
~\\
$^1${\small INAF - Astronomical Observatory of Roma, via Frascati 33, 
    I-00040 Monte Porzio Catone, Italy}\\
$^2${\small INAF - Astophysical Observatory of Catania, Via S. Sofia 78, 
        I-95123 Catania, Italy}\\
$^3${\small INAF - Astronomical Observatory of Padova, 
    Vicolo dell'Osservatorio 5, I-35122 Padova, Italy}\\
$^4${\small INAF - Astrophysical Observatory of Arcetri, Largo Enrico Fermi 5, 
    I-50125 Firenze, Italy}\\
$^5${\small INAF - Astronomical Observatory of Trieste, via G.B. Tiepolo 11, 
        I-34143 Trieste, Italy}\\
$^6${\small INAF-Istituto di Radioastronomia, via Gobetti 101, I-40129
    Bologna, Italy}\\
$^7${\small Dipartimento di Astronomia dell'Universit\`a di Trieste, 
  via Tiepolo 11, I-34131 Trieste, Italy}\\
$^8${\small INFN - National Institute for Nuclear Physics, Trieste, Italy}\\
$^9${\small INAF - Astronomical Observatory of Bologna, via Ranzani 1, 
    I-40127 Bologna, Italy}\\
$^{10}${\small INAF - Astronomical Observatory of Capodimonte, via Moiariello 16, 
    I-80131 Napoli, Italy}\\
$^{11}${\small CINECA, High Performance System Division, Casalecchio di
    Reno–Bologna, Italy}\\ 
$^{12}${\small INAF - Astronomical Observatory of Teramo, via M. Maggini, 
    I-64100, Teramo, Italy}\\
$^{13}${\small INAF - Astronomical Observatory of Torino, Strada Osservatorio
    20, I-10025, Pino Torinese (TO), Italy}\\
$^{14}${\small Dipartimento di Astronomia dell'Universit\`a di Bologna,
  via Ranzani 1, I-40127 Bologna, Italy}\\
$^{15}${\small INAF - Astronomical Observatory of Brera, via E. Bianchi 46,
    I-23807 Merate (LC), Italy}\\ 
$^{16}${\small INAF - IASF-Roma, Via Fosso del Cavaliere 100, I-00133 Roma,
    Italy}\\ 
$^{17}${\small INAF - IASF-Milano, Via Bassini 15, I-20133 Milano, Italy}\\
$^{18}${\small INAF-IFSI, Via Fosso del Cavaliere 100, I-00133 Roma, Italy}\\
$^{19}${\small Dipartimento di Astronomia dell'Universit\`a di Padova, vicolo
    Osservatorio, 3, I-35122 Padova, Italy}\\
$^{20}${\small Dipartimento di Astronomia e Scienza dello Spazio, Universit\`a
    degli Studi di Firenze, Largo E. Fermi 2, I-50125 Firenze, Italy}\\
$^{21}${\small Universit\`a degli Studi dell'Insubria, Via Ravasi 2, I-21100
    Varese, Italy}\\ 
$^{22}${\small Universit\`a degli Studi di Milano, Via Celoria 16, I-20133
    Milano, Italy}\\
$^{23}${\small Consorzio Interuniversitario per la Fisica Spaziale, Villa
    Gualino, Viale Settimio Severo 63, I-10133 Torino, Italy} 
}

\begin{document}

\begin{abstract}
On May 27th 2010, the Italian astronomical community learned with concern that
the National Institute for Astrophysics (INAF) was going to be suppressed, and
that its employees were going to be transferred to the National Research
Council (CNR). It was not clear if this applied to all employees (i.e. also to
researchers hired on short-term contracts), and how this was going to happen in
practice. In this letter, we give a brief historical overview of INAF and
present a short chronicle of the few eventful days that followed. Starting from
this example, we then comment on the current situation and prospects of
astronomical research in Italy.

\end{abstract}

\section{What is INAF - a historical perspective}

The National Institute for Astrophysics (INAF) was created in 2001 from the
merging of 12 Astronomical Observatories (which were previously independent
entities, with their own administrations). INAF was conceived with the idea of
having one Institute that would plan and coordinate all national astronomical
research, as well as promote and manage the Italian participation into European
and International projects. The first president of INAF was Prof. G. Setti.

In 2003, a decree authored by the Minister L. Moratti led to a profound
transformation of INAF, with the inclusion of other seven Space and
Astrophysics Science Institutes of the Italian National Council for Research
(CNR) [1]. Minister L. Moratti also nominated Prof. Benvenuti as
``Commissioner'' of INAF. He was later appointed as INAF president, and resigned
at the beginning of 2007. A new officer, Prof. S. De Julio, acted as INAF
``Special Commissioner'' from May 2007 to February 2008, when the then Minister
F. Mussi nominated the current president of INAF, Prof. T. Maccacaro.

The merging process between INAF and the seven CNR Institutes mentioned above
was completed only in 2009, with the definition of the legal status of all INAF
employees. In December of the same year, a new decree concerning a reform of
all research institutes was published [2]. The aim was to guarantee a more
productive and efficient use of the public funds assigned to research. Five
experts were nominated (in April 2010) by the Ministry of Education and
Research (Miur) to work together with the current Administration
Council\footnote{http://www.inaf.it/struttura-organizzativa/cda/cda.htm} on the
new reform of INAF. The entire astronomical community (including astronomers
working at Italian Universities) contributed with comments and practical
suggestions on how to make the current organization more flexible and
efficient.

In practice, since its very foundation, {\it INAF has never reached a stable
  operational state, and it is still waiting for a definitive internal
  regulation}.

Currently, INAF comprises twenty research institutes, more than one thousand
permanent employees, of which about 600 are researchers, and a bit less than
400 researchers working on short-term
contracts\footnote{http://www.ced.inaf.it/anagrafica/}. About 500 researchers
working at Italian Universities or other institutes are associated with INAF,
and collaborate actively with INAF staff members. All previous reforms, at
alleged zero cost, but actually reducing the available resources, have caused a
{\it long period of adjustment, entrenched with organizational and economical
  difficulties.}  In 2009, INAF has received from the Miur a budget of about 91
millions EUR which, normalised to the number of employees, is one of the lowest
among similar research institutes [3,4]. About 89 per cent of this budget goes
into salaries and administrative expenses\footnote{This percentage keeps rising
  due to the progressive reduction of funds allocated by Miur to the
  Institute.} [5]. A large part of the remaining 10 millions EUR go into the
operation of the Telescopio Nazionale Galileo (TNG - in Canary Islands), the
participation to the Large Binocular Telescope (LBT), and in the construction
of the Sardinia Radio Telescope (SRT). All these enterprises started before
INAF was established. This leaves very little (approximately 1.5 millions in
2009 - the equivalent of one ERC grant for about 1500 people!) to fund
scientific projects, to allow researchers to travel and disseminate their
scientific results and/or working with collaborators, and to invite
collaborators. It is worth noting that many of the projects in which INAF is
involved have been significantly reduced already (e.g. TNG has been closed for
a few months, and starting from 2010 INAF will not support supercomputing
facilities). It is clear that there is very little money INAF can save.

Despite progressively reduced funds and a tormented history, INAF has been (and
remains, but see below) an institution of great relevance to Italian science
and technology. It occupies a prominent position in research, both at national
and international levels, and is involved in a number of outreach and didactic
activities\footnote{http://193.206.241.5/struttura-organizzativa/dsr$\_$1/dsr}.
According to the most recent evaluation of research activities in Italy, INAF
is the first among the Italian research institutes in the field of physical
sciences [6]. An analysis of the scientific productivity carried out by
independent agencies (ISI Thompson) shows that Italy is fifth worldwide for
activities in the field of astrophysics ($10.3$ per cent of the world
productivity!). Out of the eighty-six Italian researchers who are among the
most cited in the world, thirteen work in the field of astronomy, and are
either working at or associated with
INAF\footnote{http://hcr3.isiknowledge.com/home.cgi}. These parameters have
remained more or less constant in the past ten years, thanks to past
investments and to the quality of Italian researchers who have partially offset
the very low funding level by securing independent research grants and
contracts (e.g. from the Italian Space Agency and also from the European
Community). Despite the efforts of Italian researchers, a first negative
signal concerning the scientific productivity was measured in 2008, and had
been anticipated by many reports of external Visiting Committees to the INAF
institutes [7]. E.g. from the Report on OA-Trieste: {\it Vital scientific
activities [\ldots] are literally being strangulated at AOTS by the current
level of funding. The VC believes that AOTS is very likely to undergo a rapid
decline of its current scientific and technological level, even of its
capability to operate as an effective research institution in the next few
years, if the funding situation is not improved}. From the Report on IASF-MI
and OA-Brera: {\it Realizing that such an inappropriate allocation of funds to
Institutes which undoubtedly score amongst the very top of Italian Astronomy is
a consequence of an inadequate funding of INAF as a whole \ldots}.

\section{May 27th 2010 - Italy turns off the stars}

On May 27th 2010, the Italian astronomical community learns with great concern
that the government plans to dismantle INAF and suddenly merge it into the CNR,
as part of an economic measure made necessary by the recent financial
crisis. The community fears that such a decision would have dramatic
consequences on the Italian astronomical community, jeopardising the relevant
contribution that INAF provides to major European and international projects
(e.g. LBT, ALMA, SKA, ELT), and effectively destroying international
agreements, with severe consequences for scientific, technological and
industrial activities in Italy. INAF would lose its scientific independence to
end up in a ``cauldron'' of large institutes. All this without leading to any
appreciable money savings, and without any clear scientific and administrative
plan on how this new merging should happen in practice.

As stated in an open letter by the Scientific Advisory Committee of INAF to the
President of the Italian Republic Giorgio Napolitano {\it the government
  decision sounds like a mockery, in addition to being a tragic mistake}.
Ironically, the very same decree discusses about favouring the return of {\it
  brilliant Italian brains} working abroad. Why should these `brains' come back
to a country where there are no positive perspectives for their future?

The community is astonished, shocked by the news (most of the Italian
astronomers - including the INAF president - learned this from the newspapers)
that INAF is listed among several institutes considered {\it useless} [8,9]. An
April Fools' Day prank? No, it is not \ldots Messages rapidly overflow all our
mailboxes. Is Italy turning off the stars?  Is the homeland of Galileo leaving
his heirs orphans? What is going to happen to us? And to all young (and
not-so-young anymore) people working on short-term contracts?  And to our
projects funded on European or International Grants?  Several protests take
place in Italy, and not just from researchers working at INAF. Several other
institutes have gone under the same hatchet.

\section{May 31st 2010 - Decision postponed?}

On May 31th 2010, the draft of the decree is read and signed by President
Giorgio Napolitano. The final version of the decree, that is currently
waiting to be turned into `law', does not include INAF among the institutes to
be suppressed and/or incorporated in other institutes.

INAF is safe. For now. There is no reason to rejoice though.

It was clear from reading the preliminary draft of the decree already, that
even if INAF was saved, life as a researcher in Italy would become more and
more difficult. The cuts to the investment in research are
conspicuous. Perspectives for new opening positions are scarce: the current
measure foresees a reduction of 80 per cent in the `turnover', i.e. for every
ten people retiring, only two new positions will be opened.

How is INAF going to survive with a severely reduced budget?\footnote{At
  present, there are different interpretations of the economic measure, but
  they all agree that INAF (and not only INAF) will have negligible funding for
  research activities.} All programmed cuts are likely going to bring to a very
significant reduction of short-term contracts and trigger a new brain drain of
young (and also not so young) researchers to foreign countries. At the same
time, it will become very hard to attract foreign researchers to our
country. {\it This implies that Italian astronomy will lose the most productive
  and competitive part of their researchers}. Not the best way to save money,
at least not on the long term. The prediction is that of a rapid and
irreversible decline of Italian astrophysical research.

The cuts are not just for INAF and astronomical research. They represent a
constant of the past few years in Italy, and risk to strike the final blow to
Italian public university teaching and research [10,11].

\section{Adopt an Italian Astronomer}

The rapidly worsening conditions and frightening perspectives have led us to
the following (provocative) initiative:
http://adoptitaastronom.altervista.org/index.html. At this webpage, you will
find a copy of this document, and the CVs of all its authors. This is {\it what
  we do}, this is {\it who we are}. One day (sooner than later?) you might well
find all these CVs among the application material you will receive.

In the meantime, aware that our astronomical competences risk to be lost,
we propose ourselves for a series of lessons/seminars at your Institutes, so as
to {\it plant a seed of knowledge that was born and grew up in our country}. If
you wish to give us your support by inviting us to your Institute, please send
an e-mail to the address: adoptanitalianastronomer@gmail.com, and help us to
circulate this letter within the international astronomical community. We plan
to make all seminars and lectures that will be given in the framework of this
initiative publicly available.

\section{Conclusions}

These are very difficult times. 80 million citizens in Europe currently live
below the relative poverty line according to EU statistics (10 per cent of the
population in Italy) [12].  Explaining how this happened goes beyond the aims
of this letter (and the competences of its authors).  It is, however, our
opinion that the budget currently invested in science and research is {\it not}
among the causes of the current situation, in Italy or anywhere else. Instead,
it is our firm belief (which seems to be shared by most countries\footnote{See
  e.g. http://ec.europa.eu/invest-in-research/index$\_$en.htm. The Lisbon
  Treaty, entered into force on 1 December 2009, states that {\it The Union
    shall have the objective of strengthening its scientific and technological
    bases by achieving a European research area in which researchers,
    scientific knowledge and technology circulate freely, and encourage it to
    become more competitive, including in its industry, while promoting all the
    research activities deemed necessary by virtue of other Chapters of the
    Treaties}. Italy has signed this Treaty.}) that investing in the future of
education and research is a top priority, and that this becomes even more
important in times of economic crisis\footnote{See also
  http://www.bmbf.de/en/96.php}. Research and education {\it do have} mid and
long-term positive effects on the economical, technological and industrial
situation of a country. {\it Curiosity-driven} research has itself an important
value, being perhaps the only real driver for significant and long term
progress (also in economical terms: e.g. the web has been invented by particle
physicists!).  Cutting on education and research means cutting on the
future. As scientists, before than as Italians, we cannot but express our
dismay and deep worry that the decisions currently being taken are inevitably
leading to a further cultural and economical impoverishment of our country.

The usefulness of basic, intermediate, and high education, as well as of
research, is being questioned in Italy, mainly because they do not produce {\it
  quick} and {\it easy} money. We realize that, unfortunately, this approach
and way of thinking is not limited to our country (although Italy probably
represents a dire example of cultural decay). As scientists, we cannot turn
aside from fighting this way of thinking, that we consider {\it blind, foul, and
irresponsible}.  It is {\it not} these principles that we want to hand on to
future generations of astronomers, Italians, and citizens of this world.

\section{Acknowledgements}
We thank all our foreign collaborators, for having always trusted and
appreciated the carefulness, dedication, and passion driving our research, as
well as its social and economical value.  We apologize to anyone who might not
find that the scientific content of this paper is enough for it to appear on
the arXiv.

\section{References}

[1] http://193.206.241.5/presentazione/normativa/dl$\_$138

[2] http://wwwrntta.mi.infn.it/Documenti/leggi-e-decreti/DecretoRiordinoEPR.pdf

[3] http://www.lascienzainrete.it/node/1489

[4] http://rasta.media.inaf.it/data/rasta-2010-06-22-49077.pdf

[5] http://193.206.241.5/struttura-organizzativa/presidenza/piano-triennale/\\

\vspace{-0.8cm}\hspace{0.6cm}pt-2009-2011$\_$files/pt-2009-2011.pdf

[6] http://vtr2006.cineca.it/php5/relazione$\_$civr/output/totale.pdf

[7] http://193.206.241.5/struttura-organizzativa/presidenza/piano-triennale/\\

\vspace{-0.8cm}\hspace{0.6cm}pt-2009-2011$\_$files/A4\%20-\%20testo\%20copia.pdf

[8] http://www.blitzquotidiano.it/politica-italiana/\\

\vspace{-0.8cm}\hspace{0.6cm}ecco-i-27-enti-inutili-che-la-manovra-finanziaria-sopprimera-397874/ 

[9] http://beffatotale.blogspot.com/2010/05/enti-inutili.html

[10] http://www.repubblica.it/scuola/2010/06/24/news/\\

\vspace{-0.8cm}\hspace{0.6cm}presidente$\_$crui-5110009/?ref=HREC2-7

[11] http://www.nature.com/news/2010/300610/full/466016b.html

[12] http://www.2010againstpoverty.eu/about/?langid=en

\end{document}